\documentclass[runningheads]{llncs}
\usepackage{graphicx}
\begin{document}
\title{Securing Manufacturing Intelligence for the Industrial Internet of Things\thanks{Supported by the Centre for Industrial Analytics, University of Huddersfield.}}
%
%
\author{Hussain Al-Aqrabi\orcidID{0000-0003-1920-7418}and
Richard Hill\orcidID{0000-0003-0105-7730}\and
Phil Lane\and
Hamza Aagela}
\authorrunning{H. Al-Aqrabi et al.}
\institute{Department of Computer Science, University of Huddersfield, Huddersfield, UK
\email{\{h.al-aqrabi,r.hill,p.lane,hamza.aagela\}@hud.ac.uk}}
\maketitle              
\begin{abstract}
Widespread interest in the emerging area of predictive analytics is driving industries such as manufacturing to explore new approaches to the collection and management of data provided from Industrial Internet of Things (IIoT) devices. Often, analytics processing for Business Intelligence (BI) is an intensive task, and it also presents both an opportunity for competitive advantage as well as a security vulnerability in terms of the potential for losing Intellectual Property (IP). This article explores two approaches to securing BI in the manufacturing domain. Simulation results indicate that a Unified Threat Management (UTM) model is simpler to maintain and has less potential vulnerabilities than a distributed security model. Conversely, a distributed model of security out-performs the UTM model and offers more scope for the use of existing hardware resources. In conclusion, a hybrid security model is proposed where security controls are segregated into a multi-cloud architecture.
\end{abstract}
\section{Introduction}
The Internet of Things (IoT)\cite{ashton2009,gubbi2013,chen2015,gartner2015,ikram2015,sundmaeker2010} and the Industrial Internet of Things (IIoT)\cite{pwc2016,wang2016a,wang2016b,liu2014} is a collection of enabling technologies that can provide data-driven visibility of physical systems. Industrial organizations, particularly those in the manufacturing industry, make use of myriad material handling and conversion processes that require data to control at both micro and macro levels. At the micro level, data is an intrinsic part of a control loop to assure quality and repeatability of the process. At the macro level, data is used to coordinate the logistics of material movements between individual processes, and manufacturing units, for local, national and even international distribution.

As the costs associated with technology constantly reduce, and the availability of hardware platforms and Software-as-a-Service (SaaS) becomes more commonplace, more and more enterprises are installing IIoT devices to increase the awareness of their operations\cite{hill2017}.

The installation of new manufacturing plant often includes the capability to extract data about operations in real time. However, older plant does not have such capabilities, and from a Return On Investment (ROI) perspective, there is no justification to replace plant that is functioning satisfactorily. Thus, there is scope for the retro-fitting of IIoT devices to plant and machinery, in order to gain valuable insight into any potential process optimisations that might occur as a result of having improved visibility of process data.

Once there is an IIoT infrastructure in place, where the process data can be monitored, collected and analysed, there is range of opportunities to visualise process data, as well as the macro level whereby collections, or even entire supply chains can be monitored.

At the lowest level of observation, raw sensor data is collated; examples of such data include: rotational speed; current draw; temperature; pressure; humidity, and other readings of physical properties that can be converted into electrical signals via transducers.

One phenomenon that becomes apparent with the installation of IIoT is the considerable increase in volume of fast-moving data that has to be captured, transported, stored and managed. Ultimately, to realise the full potential of such data requires analysis. The addition of modelling and forecasting operations, using the data, gives rise to the emerging interest in the field of \emph{predictive analytics}.

To date, the response from enterprise software system vendors has been to advocate high-bandwidth network upgrades, in conjunction with utility computing solutions such as cloud computing. Large organizations, who possess their own data centres, may adopt a private or hybrid cloud approach, so that they can increase the utilisation of resources for on-demand computational requirements.

Enterprises without such resources are then faced with the decision of whether to acquire additional hardware via capital expenditure (CapEx), or to make use of cloud computing (perhaps via Platform, Infrastructure or Software as a Service) which is accounted for as an operational expenditure (OpEx).

Invariably, there is insufficient justification for enterprises to make significant CapEx investments, especially when the same capabilities can be provided as a service, on demand.

The adoption of analytics platforms from a remote cloud service does mean however, that process data from the enterprise must be transported outside of the boundaries of the organization, a prospect that is viewed with some skepticism.

There are two issues at play here. First, organizations wish to retain full control of their data, and this can only be assured by keeping data on the premises. 

Second, whilst an enterprise may yet to have yielded any value from the analysis of the raw data, it is conceivable that the analysed sensor data, stored in a cloud which is off the premises, is an obvious target for attack and therefore this introduces a vulnerability into the security of the organization's operations\cite{katic2006,rosenthal2000,kadan2012,fernandez2007,ahmad1,brankovic2000,alaqrabi2012}. Raw data that has been processed, analysed and used to construct models of the business, is the key source of knowledge for \emph{Business Intelligence}.

From an organization's perspective, this data will contain valuable insight into the detail of how raw materials are converted into products. Traditionally this knowledge has been retained tacitly by human machine operators and managers, but the adoption of IIoT equipment to sense and gather this data means that valuable Intellectual Property (IP), that is the core of any organization's competitive advantage, can be managed just like any other asset.

The act of collating knowledge brings significant possibilities into being\cite{blanco2009,stobla}, whilst also increasing the risks of the knowledge being lost or stolen\cite{abadi2009,fienberg2006}. Whilst a malicious employee might share the `secrets' of a process that they manage, it would be difficult to envisage how a larger security breach could be coordinated. Conversely, a malicious attempt to attack a remote server might provide a back door into a system that can be copied, disrupted, or even monitored covertly.

Whilst one of the benefits of using a remote cloud service is that the levels of security are of a higher grade than most small enterprises can afford, there remains a widespread reluctance to trust systems that are remote with the core IP of an industrial enterprise. 

In the light of these challenges, this work explores how the security of IIoT-centric BI might be augmented to increase trust between physical plant within an organization's premises, with the remote analytics capability that is hosted remotely.

In particular, we propose, simulate and evaluate two models, that offer significant improvements in security. These improvements will be of interest to enterprises who cannot yet justify the wholesale management of IIoT analytics management on premises, by mitigating some of the additional risks presented by established off-premises solutions.

\section{Business Intelligence services}

The infrastructure of BI is a complex collection of functions that inter-operate to consume, share and process data for reporting purposes. Data sources are varied and include operations data from enterprise systems and sales order processing, as well as process data for the coordination of operations activities \cite{alaqrabi2014}.

The adoption of IIoT gives enterprises the ability to include in-process data from manufacturing operations, increasing the scope and depth of business reporting that is possible. In general, transaction data is stored in relational databases, data-marts and data warehouses, which is available for query by the BI system \cite{alaqrabi2013}.

As such, BI is an enhanced interface that facilitates human interaction with the organizations's data, leading to judgements, decisions and actions around operational interventions. This important role of BI elevates its status beyond that of traditional IT reporting infrastructure, and thus creates a greater dependency between its users and the business.

Such is the dynamic nature of business, especially in the fast-moving manufacturing industry, that the requirements placed upon data usage will invariably change depending upon the needs of the business, whether it be from the external environment, internal disruptions within the physical systems, or the aspirations of the business executive.

As business requirements change, there will be a corresponding change in the rules governing data acquisition, transformation, formatting, and finally, loading.

Each of these transactions and human interactions are opportunities for vulnerabilities either through human error or nefariously. Attempts to model complex, multi-agency systems\cite{beer2003} have enabled greater understanding and communication of complex human interactions with sensitive personal, in this case health data. As a consequence, a corpus of security challenges lie in the procedural and human interactions with the BI system\cite{alaqrabi2013}. The work described in this article is focused on the technical security challenges and solutions when BI is hosted off the premises.

\subsection{BI architecture}

BI processes are often composed from orchestrated, collaborating services, which are consumed by different users. In the broader domain of businesses that collect and consume data from IIoT systems, and who have opted to utilise BI as a service, each of the corresponding component services that constitute the overall BI application may, in fact, request aggregated data from multiple cloud systems, each of which may be in a different security realm.

These internal services have to be executed dynamically at runtime, and therefore the requisite authorisation and permissions need to be in place to allow such actions to happen.

Cloud-based systems are inherently heterogeneous in nature which necessitates the automation of authorisation wherever possible, to facilitate the timely and seamless delivery of BI services. It is a considerable technical challenge to enable this secure collaboration.

In order to address this security challenge, an authentication framework is required to establish trust amongst BI service instances and users by distributing a common session secret to all participants of a session. We have addressed this challenge by designing and implementing a secure multiparty authentication framework for dynamic interaction, for the scenario where members of different security realms express a need to access orchestrated services\cite{alaqrabi2012}.

This framework exploits the relationship of trust between session members in different security realms, to enable a user to obtain security credentials that access cloud resources in a remote realm. The mechanism assists cloud session users to authenticate their session membership, thereby improving the performance of authentication processes within multiparty sessions.

We see applicability of this framework beyond multiple cloud infrastructure, to that of any scenario where multiple security realms have the potential to exist, such as the emerging IIoT within the manufacturing industry, commonly referred to as an essential component of Industry 4.0\cite{pwc2016,wang2016a,wang2016b,hill2017}.
\subsection{BI Security challenges and controls}
As a business builds its BI capability, the complexity of new BI workflows increases as workers begin to ask more insightful queries of their data. As such, it can be very challenging for information and security architects to be able to design the necessary safeguards that will enable business users to pose queries, without hindering the potential for more complicated inquiry at a future date.

The most comprehensive BI framework requires full ownership of processes that can access data at multiple level of granularity. Data and information that is useful exists within all aspects of an enterprise system, including, but not limited to: hardware and network systems; data marts and warehouses together with associated metadata repositories; OLAP servers; the data presentation and application services layers, as well as the appropriate layers of authentication\cite{katic2006,rosenthal2000,kadan2012}. In addition, the wholesale system augmentation of IIoT devices adds further complexity in terms of data volume and velocity, increased demand upon authentication services, and myriad protocols and the necessary query brokering and routing services.

Using the basic principle that data is secure by default and only shared on a `need-to-know' basis, the authentication services apply to each generator or collector of data within the system. It follows that the security controls are bound to the repositories that hold both temporal and permanent data\cite{farhan2012}.

Extract, Transform and Load (ETL) functions within data marts require this data in order to perform the correct operations on the pertinent data at the correct time. The security controls apply both to system level and human stakeholders, depending upon the nature of the interaction with the business system. Some of the data will be editable, whereas other data will be permanently stored such as financial transactions, for instance.

Multi-dimensional models for the secure management of objects within data warehouses have been proposed\cite{fernandez2007,priebe2000,cuzzocrea2012}, particularly for OnLine Analytical Processing (OLAP) queries. In this work, there are objects that are dedicated to the preservation of controls upon different operations within a multi-dimensional model. Each object was invoked on a per-request basis for each transaction.

Using Object Security Constraint Language (OSCL) to extend the Unified Modelling Language (UML), a multi-dimensional model is described. Such is the arrangement of the data repositories, together with the tight coupling of security control objects to the repositories, that a hierarchical organization of the security objects also emerges.

Whilst the cohesive approach to integrating secure control objects into the multi-dimensional data model is effective when the data domain owner is known, instances of temporal data may not have clearly identifiable owners\cite{ahmad1}, and therefore the argument for this approach is less convincing.

During the course of querying data for a particular purpose, it is likely that a particular BI user requires access to data that it is not the owner of. One example is of data streams from shopfloor manufacturing processes, that have been fused with relevant logistics data. In this particular instance (that is very commonplace), the data that is required does not necessarily have sufficient metadata to describe the constraints under which that data can be processed or even viewed\cite{alaqrabi2014}.

In this scenario we need to ensure that the controls that apply to intermediate repositories of transient data will require greater security controls, than if the data had been extracted from its source, where the domain owner of both the owner and the requester can be ascertained through conventional authentication mechanisms\cite{brankovic2000}.

One of the trade-offs of a cohesive coupling of security objects to multidimensional arrangements of repositories, is that there is an increased cost for ETL transactions in terms of performance overhead. This is further complicated as additional levels of abstraction are considered (for other purposes), such as the rules engines required for OLAP processes, that are usually separated from underlying data warehouses\cite{blanco2009,agrawal2005}.

Additional real-world considerations with regard to the transport of data within a system, especially with the integration of IIoT devices, is the likelihood of transfer over wireless networks or hybrid/public clouds, where the underlying infrastructure is not owned nor controlled by the organization. In such cases the data will be encrypted after extraction, as part of the transformation process, prior to being loaded into a repository.

So far, the discussion has explored the difficulties of implementing secure BI workflows within manufacturing information systems. Whilst it is possible to conceive of robust security controls, their application to the inherently complicated systems that develop over time within manufacturing organizations means that there can be a significant impact upon the performance characteristics of the eventual system. This impact, to an extent, can be mitigated by the use of more computing and network resources, though this requires additional investment.

However, there comes a point at which additional expenditure on infrastructure is not feasible. This point may be reached quicker with the adoption of IIoT technologies that can rapidly overload existing network architectures with extra data transport, a significant part of which is directly incurred as a by-product of the additional authentication transactions for the additional physical objects, and the associated queries that become possible as a result\cite{shu2015,li2015,baker2012}. This is also in addition to essential infrastructure services that invariably increase as more equipment is added.

\begin{figure}[htbp]
\centering\includegraphics[width=0.8\linewidth]{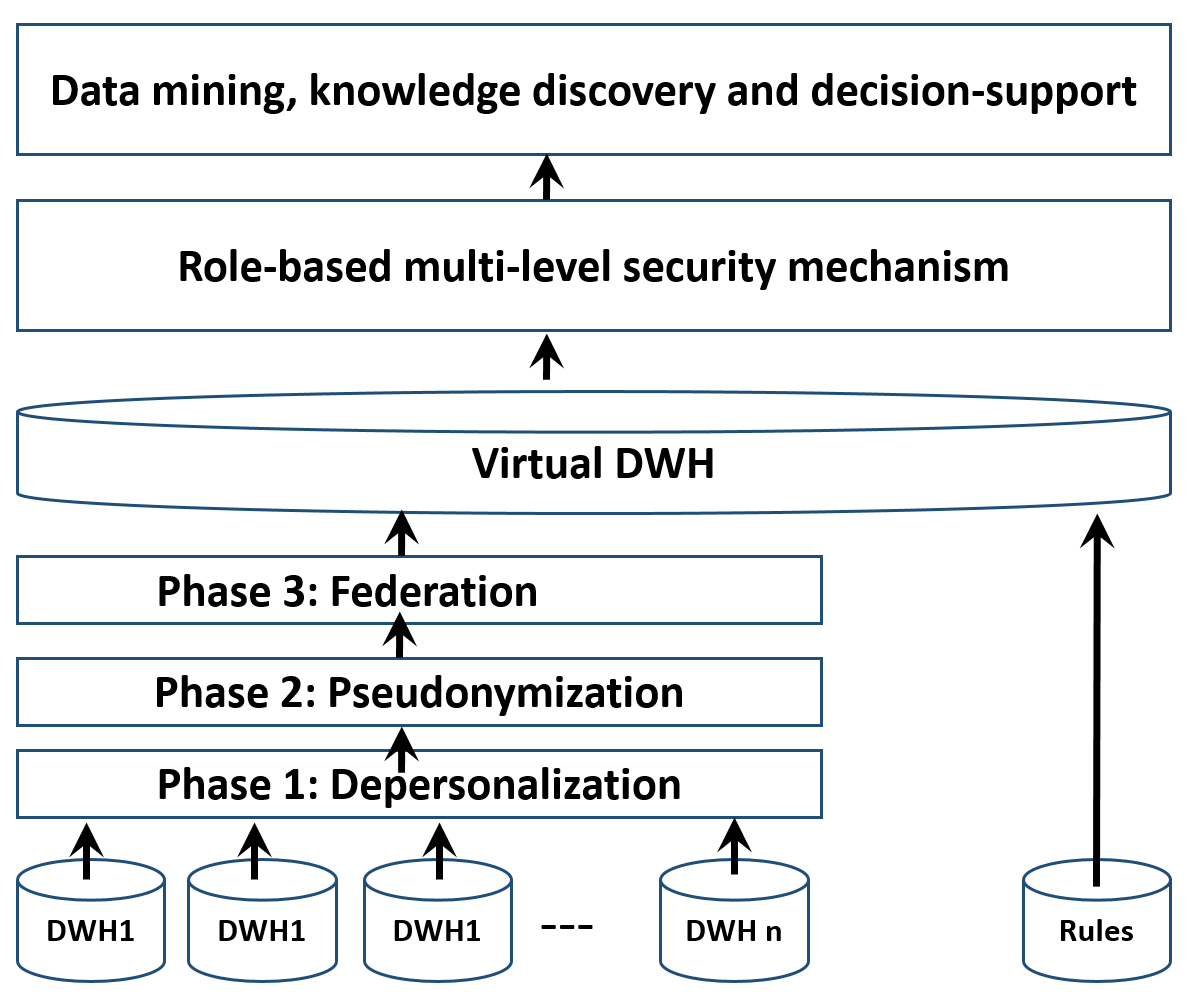}
 \caption{Distributed security for a system of federated data warehouses\cite{stobla}.} 
 \label{feddata}
\end{figure}

Figure \ref{feddata} illustrates one attempt to cope with this challenge by federating clusters of data warehouses. Each data warehouse is separated by a network switch and associated data storage, and enables data within multiple warehouses to be striped or \emph{depersonalised}, and queried as one object\cite{stobla}.
The framework \cite{stobla} assigns federated repositories to separate groups of administrators. The process of striping the data enables an element of obfuscation in that any data within one resource is meaningless without its relevant contextual information. Such an approach lends itself to utility computing infrastructure such as clouds, where the elasticity of resources provides in-demand scalability.

It follows that the striped deployment of relational databases could similarly possess elastic characteristics, and therefore any additional objects, such as for the implementation of security controls at object-level, could be accommodated in an architecture that scales when necessary\cite{abadi2009}.

Since this scalability can only occur where there is a proportion of redundant infrastructure, there is a compelling case for the adoption of cloud computing, since an organization would have on-demand access to resources that it would otherwise have to purchase and maintain itself if hosted on the premises.
\subsection{OLAP security}
OLAP is a prevalent part of BI implementations, and it provides a means by which reporting can be performed against a multi-dimensional repository.
Views of the data, referred to as `cubes', enable visualization applications such as dashboards to present data to business users. In some cases, the BI dashboards permit users to interact with the data cubes, either as a means to access different presentations of the data, or even to re-query the repositories to return new insight from the underlying data stores.

An OLAP presentation contains at least a services and a user's cube, each of which has associated with it a set of security controls that protect the cubes from unauthorized access to data\cite{wang2004}.

\begin{figure}[htb]
\centering\includegraphics[width=0.9\linewidth]{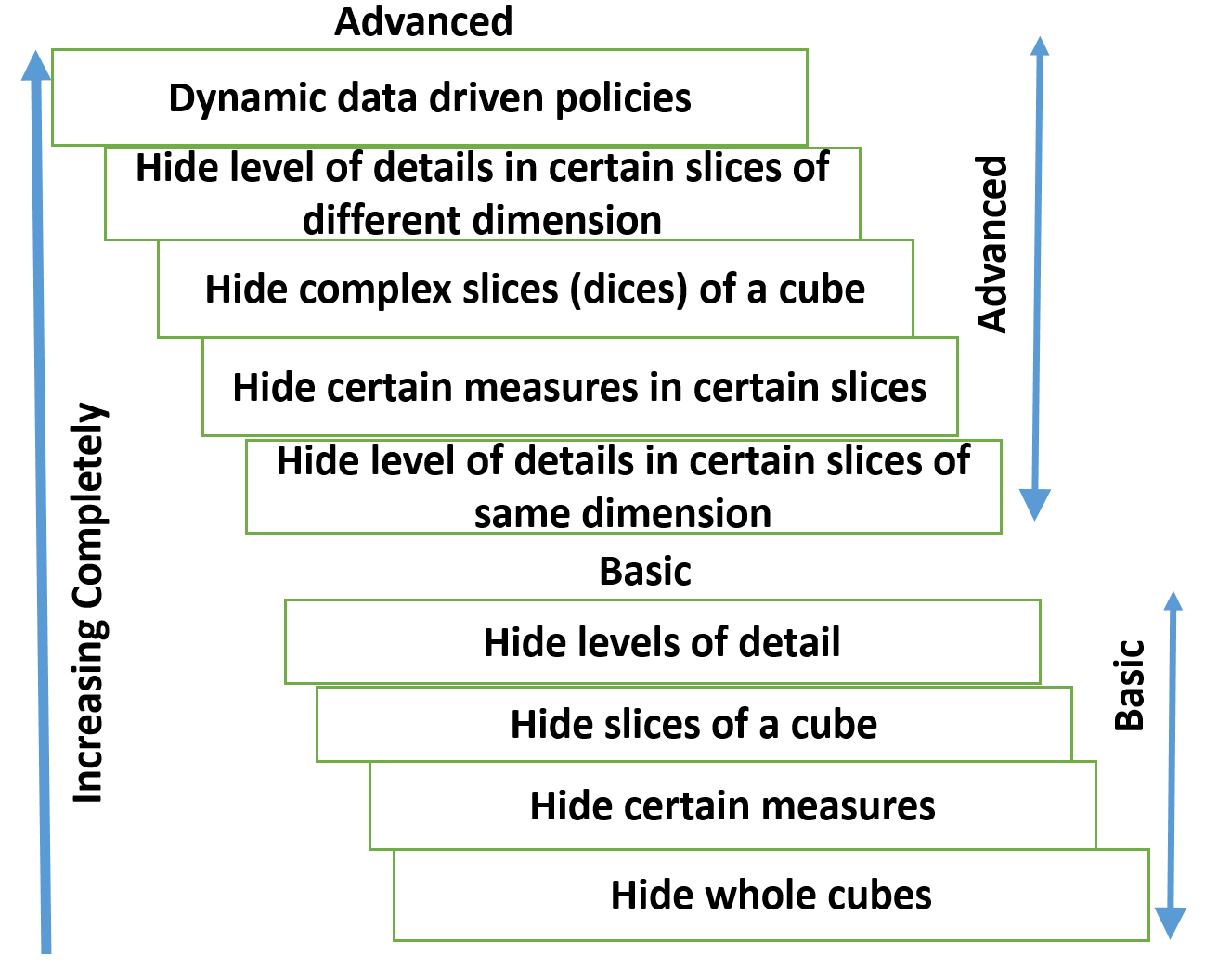}
 \caption{Distributed security controls in an OLAP cube comprising viewing and operations restrictions.}
 \label{distolapsecurity}
\end{figure}

OLAP functionality permits more complex operations to be performed on data, in a controlled way, than would normally not be permissible, nor practical with a traditional management information system\cite{priebe2000}. The granular control of this extended functionality is administered by a series of OLAP operations, themselves being constrained by a Multi-Dimensional Security Constraint Language (MDSCL)\cite{alaqrabi2013}. Figure \ref{distolapsecurity} illustrates the security controls to restrict view operations.
As the business users create queries, via the BI architectural concept of a cube, the system associates the appropriate logical access and operation execution controls as befits the outcomes that the user desires. These controls form part of a library of controls that can then be re-purposed to apply to future cubes as new queries are generated.

Business users can then access the controls via instances of cubes, typically as part of a query modelling process carried out by the users \cite{priebe2000} or system administrators.

Since the OLAP organization facilitates the creation and customization of different views of the data, it is important to ensure that inadvertent access is not granted to unauthorized users, in order that the privacy of both users, users' access to data, and the data itself is preserved.

The Secure Distributed-OLAP aggregation Protocol (SDO) is a means by which such privacy can be assured. Using SDO, the owners of views of multi-dimensional queries can be secured and logged. This enables the control and prevention of unauthorized access as described above, and manages the data that is held about a user, their permissions and the details about their interactions with data objects\cite{cuzzocrea2012}.

These `privacy metrics'\cite{wang2004} will be hierarchical in structure \cite{agrawal2005}, and reflect detailed signatures of users' interactions, in relation to the sensitivity of the data, as well as the originating privacy criteria of the repositories\cite{wang2004}, through their interactions with the various cubes. The result of this is that details of the users are then stored in a log for future analysis.

A variation of this proposal\cite{fienberg2006} provides the introduction of further controls (and overhead) to manage constraints on the OLAP cubes and underlying repositories. By the augmentation of Intrusion Detection and Prevention Systems (IDPS) into the scheme, requests made to repositories can be marshalled as an additional layer of protection against unauthorized access.

Similarly, Lightweight Directory Access Protocol (LDAP) is a means by which access to services on a network can be managed through an authentication mechanism. At a lower level, Secure Socket Layer (SSL) protocols can also be utilised to ensure that interactions within objects are transacted via an encrypted exchange\cite{fienberg2006}.

This has been a considerable challenge for the research community for some time now. There appears to be a preference for security mechanisms that are distributed across systems, with individualised controls at object level. In contrast, the Unified Threat Management framework (UTM) approach considers the system as a whole entity that must be protected as such and requires centralized governance.

Conceptually a distributed approach is more resilient to nefarious attacks; however the practicality of managing such a system is far more challenging than a centralized governance model. In addition, there is the potential to optimize and streamline the effect of controls upon system resources when it is centrally administered, whereas the more flexible, object-level security controls, present a potentially greater load upon network infrastructure through the greater volume of authentication activity for a given task.

However, the emergence of IIoT, and the awareness that systems need to be designed and built to scale elastically in the future, is also generating enthusiasm for cloud-provisioned resources and approaches to software development such as Microservices Architectures (MA)\cite{shadija2017}. We have attempted to approach the modelling of this situation by recognising that a manufacturing organization will want to adopt IIoT devices in a fluid and scalable way, and have adopted good practice for the modelling of networks that may include mobile sensors and other entities as advocated by\cite{garcia2015}.

In reality, the limit of computing resource is constrained either by physical infrastructure, or via the costs associated with a utility computing provision, so there is a need to understand the implications of each of these approaches, in order for organizations to ascertain the most appropriate approach. To date, there is no evidence of such a comparison, which has given rise to the work described in this article.

We describe the modelling and simulation of both a UTM approach versus a distributed approach to security, in the context of the manufacturing domain where IIoT devices are included as part of the system. Since there is a need for rapid allocation, de-allocation and re-allocation of resources, we have adopted a cloud-centric model, though this does not in itself necessitate an off-the-premises solution.
\section{Description of the Models}
Since the UTM approach is centralized, it is logical to group security controls into the following layers: network, transport, session and presentation. For the distributed approach, application layer security is appropriate, utilizing each of the components within the infrastructure.

Both scenarios have been modelled in OPNET. Model A uses a UTM cloud within a demilitarized Zone (DMZ) to contain the security components that are required to govern the whole system. All connection requests to the BI application are marshalled via the UTM cloud.

\begin{figure}[htb]
\centering\centering\includegraphics[width=0.95\linewidth]{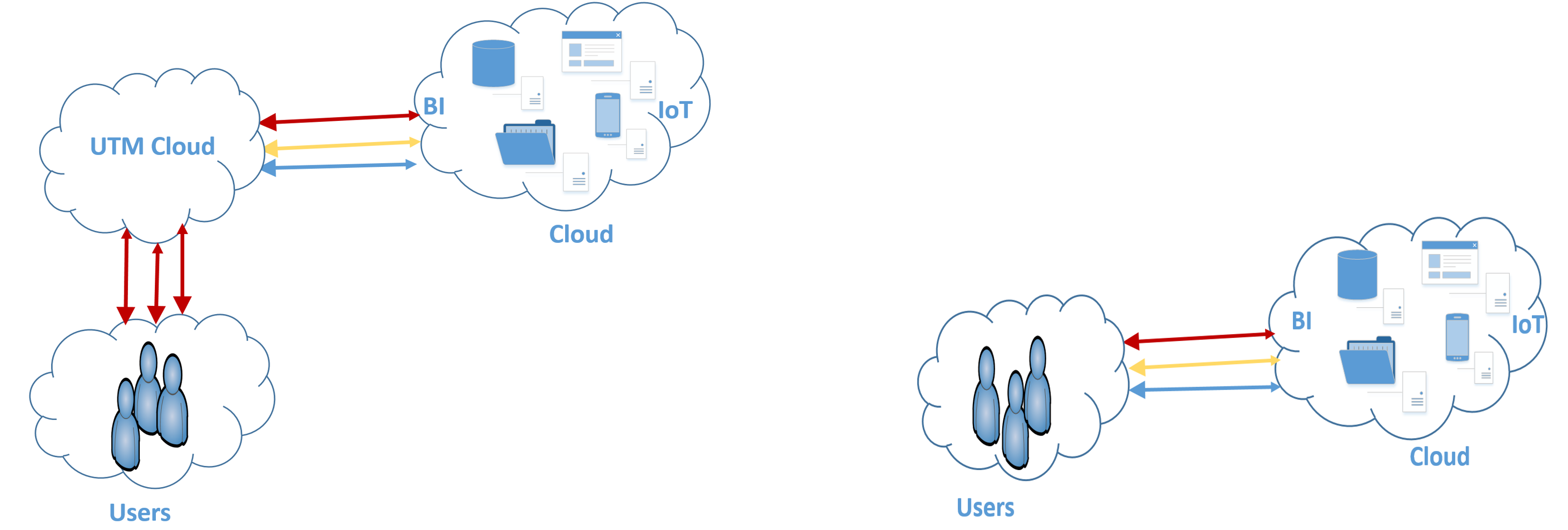}
\caption{a) Unified Threat Management (UTM) model;    b) Distributed security model.} \label{bothmodels}
\end{figure}

In contrast, Model B has no requirement for a UTM cloud as the security components have been distributed at object level, hence all connection requests are made directly to the object that is to be invoked, via appropriate object-specific security controls. Both approaches are illustrated in Figure \ref{bothmodels}.
OPNET allows the creation of bespoke application configurations, which includes the ability to assign them to different resources (in this case servers) according to the role that is played within the overall system.

For example, the {\tt OLAP\_DASHBOARDS}, {\tt DW\_DM}, and {\tt OLAP\_VIEWS} are all represented by OPNET \emph{profiles}, and are assigned within the BI application and database cloud in Model A (UTM model). The remainder of the controls reside within the UTM cloud in the DMZ.

Since the UTM cloud (and DMZ) is not present in Model B (distributed security model), the entirety of the roles are undertaken by the BI application (via OLAP) and database (in this case a data warehouse) servers within the BI cloud. Therefore, Model B has the role of security distributed across all the application and database servers, embedding the marshalling at object level. Model A's centralisation of security roles into a separate security domain contrasts with the distributed approach of Model B.
\subsection{Model A: Unified Threat Management approach}
\begin{figure}[htb]
\centering\includegraphics[width=0.95\textwidth]{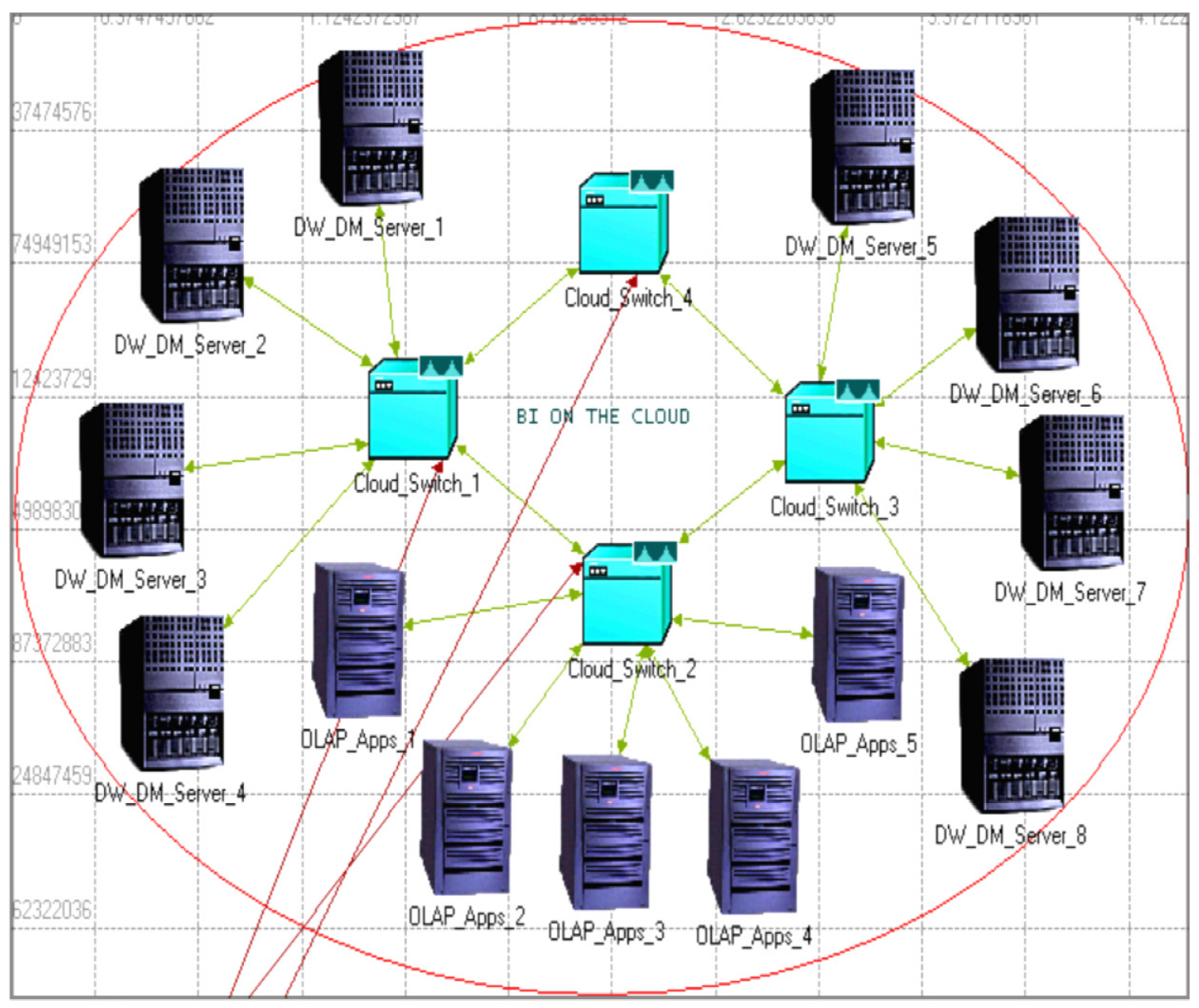}
\caption{The BI server arrays forming a cloud infrastructure.} \label{bicloud}
\end{figure}

Figure \ref{bicloud} illustrates the BI system which is composed of two server arrays. The BI on cloud (comprising of application servers and database servers) has been modelled employing four different arrays. There is an array of five OLAP servers connected to cloud switch 2, an array of four database servers connected to cloud switch 1, another array of four database servers connected to cloud switch 3 and an array of four high-end Cisco switches (Cisco 6509 routing switches). The two sets of database server arrays have been presented to model the concept of splitting records between two hardware clusters.
Each of the databases that stores the temporal and permanent data are represented by the {\tt DW\_DM} servers. OLAP applications are hosted on server arrays, including both the applications and the view cubes. Server connectivity is provided via simulated Cisco 7000 switches.
\begin{figure}[htb]
\centering\includegraphics[width=0.95\textwidth]{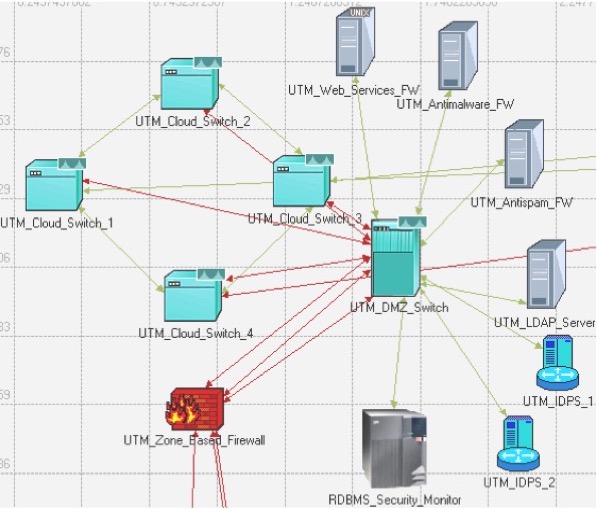}
\caption{Security components within the  Unified Threat Management cloud.} \label{utmcloud}
\end{figure}
Figure \ref{utmcloud} illustrates the security components contained within the DMZ.

Each user session is directed through a zone based firewall, which isolates all user sessions and prevents them from accessing the cloud switches directly without the appropriate authority. All user sessions are forwarded to the UTM switch.

Source and destination rules for each client and server have been created to inform the simulation as to which client and server is the target for a particular request. This feature within OPNET had been employed to create complex traffic flows between clients (representing users) and the relevant cloud servers.

The destination servers for the clients (in the extranet domain of BI users) are the servers with various security roles in this DMZ, and the cloud OLAP servers are in turn the destination servers for these DMZ based security servers. The traffic is now strictly rule-based, an approach that is an established method amongst the body of research work in this area.

The {\tt RDBMS} security monitor holds all the rules deployed for ETL processes, and others are regular security servers. The {\tt LDAP} server is deployed as a reference database of all user accounts. Hence, the {\tt RDBMS} security monitor has a destination rule pointing towards this server and this server, in turn points towards the cloud servers.
\subsection{Model B: Distributed security model}
Model B is presented in Figure \ref{bothmodels}. As described earlier, the security components are integrated with the BI servers in the cloud-based arrays, and all client requests are made directly upon the cloud switches. Since the centralized UTM cloud is absent in Model B, the necessary security components and services have been embedded within the OLAP, database and data warehouse servers as follows:
\begin{itemize}
\item {\tt BI\_DW\_DM}: the data warehouse and data marts servers;
\item {\tt BI\_Application}: OLAP apps servers;
\item {\tt BI\_Security\_UTM}: all security services are enabled on each server.
\end{itemize}
This represents the concept of distributing embedded security  controls, as advocated by the research community. The application destination preferences have been directed towards the OLAP application servers, which in turn have their destination preferences as the data warehouse and data mart servers.
\section{Results}
In this section, the simulation results of the two approach are presented, together with a discussion about the relative efficacy of each approach to security.
\subsection{Performance}
For Model A, the average response time to database queries across the entire network exceeds 20 seconds, and the average HTTP object response time can take up to 3 seconds.TCP delays are in excess of 20 seconds, with TCP segment delays of 4 seconds. The TCP retransmission count exceeded 1000 on two occasions.

%
\begin{figure}[htbp]
\centering\includegraphics[width=\textwidth]{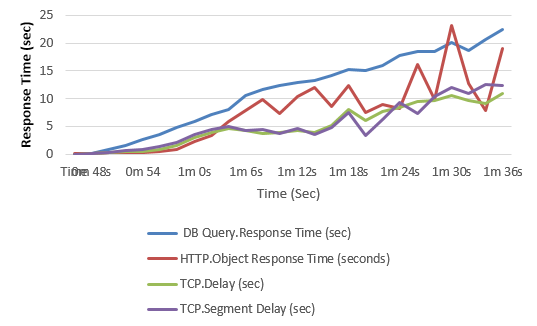}
\caption{Performance of Model B.} \label{PerfModelA}
\end{figure}
All the user traffic is being routed through the security services servers in the UTM cloud. The users are connected to a zone-based firewall, which routes their traffic to the security servers. With this configuration, there are two hops from user workstations to the BI application servers on the cloud. In this arrangement, the capacity may not be an issue, but the routing of traffic through the servers placed in the Demilitarized Zone (DMZ) is adding a delay component onto all of the servers.

\begin{figure}[htbp]
\centering\includegraphics[width=\textwidth]{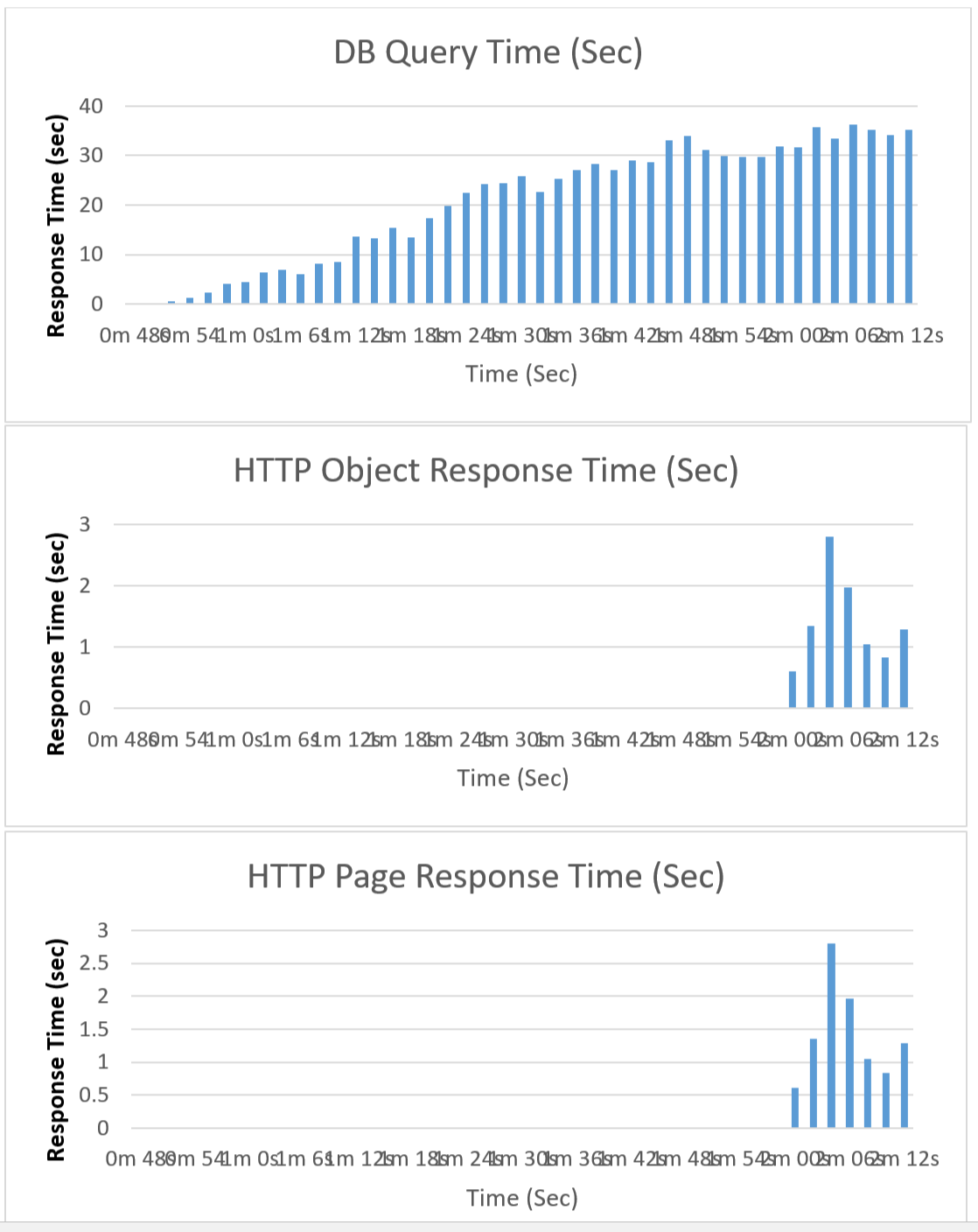}
\caption{Performance of Model B.}\label{PerfModelB}
\end{figure}

The performance report of Model B is presented in Figure \ref{PerfModelB}. In this model, the performance is within the expected limits and there are no TCP delays, TCP segment delays and TCP retransmissions. We observed that the performance of database query response and HTML page/object responses is superior to Model A and is satisfactory from a user’s perspective. The response times returned in this model are as expected from a cloud hosted BI application. Removal of the UTM cloud layer has ensured that all the session bottlenecks in the network are eliminated.
User sessions are served by the BI cloud based server arrays for application requests and for the security sessions as well. The load of the security sessions are distributed amongst all servers in the OLAP and data warehouse arrays given that the databases serving the security applications will also be partitioned and spread across the servers.

Massively parallel processing models\cite{alaqrabi2013} hold for security services as well. This is essentially a practical form of implementing the distributed embedded security components recommended by the research community as evident from the literature review.

As the trend towards service distribution and embedded computation continues, is is conceivable that a UTM cloud owner may increase the number of significantly large sized arrays of servers, and then partition all security related databases to produce a massively parallel system. In such a scenario, a UTM might be the first choice for a system architect. Innovative methods are emerging for the meta scheduling of multi-cloud instances, in a bid to counter the increased complexity of balancing resource efficiency against the need to provide an acceptable level of service\cite{sotiriadis2013,sotiriadis2015}.

However, whilst the performance at the UTM cloud will definitely improve, network and session layer congestion cannot be reduced. This is because the system will be a cascade of two large clouds, and the cascade itself will be a bottleneck. The effects of a massively parallel database query-distribution system will create difficulties when two large arrays are cascaded by the joining of two clouds.

This limitation will exist when two clouds are interconnected and the user sessions will be allowed to pass through one cloud to the server arrays of another. Performance will improve when the users are allowed to connect to two clouds independently for different purposes, and there is no inter-cloud cascade. However, such a system will not be effective from security point of view, because the user sessions need to pass through a common checkpoint before being allowed to access application resources.

We must also consider that the demands being placed upon BI applications are continuing to stretch performance as users derive new insight from innovative applications, such as automated object-tracking from video stream data\cite{yaseen2017}, which is finding many applications in the manufacturing domain.
\subsection{Security efficacy}
Model A incorporates both a UTM and DMZ, which are well proven security concepts, whereas the distributed approach of Model B is yet to be proven to the same degree. To summarise:
\begin{enumerate}
    \item For Model A, each user is marshalled to the BI application via the DMZ. In Model B, users connect directly to the BI application components.
    \item Cyber attacks into a Model A arrangement are faced with only one secure entry point to overcome. Again, for Model B there are multiple potential vulnerabilities that will increase as more components and devices are added to the system over time;
    \item Security updates and maintenance is a constant requirement that will be managed via the UTM cloud in Model A. For Model B, this results in a very challenging environment whereby the security responsibilities of components are individualized.
\end{enumerate}
However, Model A suffers from performance related issues. Close scrutiny of the simulation results shows that {\tt UTM\_DB\_ACT\_MON} is generating maximum traffic, a cause of the performance degradation.

The Security-as-a-Service (SECaaS) approach via a UMT cloud is attractive to adopt as user sessions pass through the UTM zone-based firewall, whereas user sessions directly hit the cloud servers in the distributed security model (Model B). In addition, security administration is much easier than the distributed model.

However, the issue of security vulnerability through the multiple entry points within Model B could be addressed by the use of a cloud server to prevent an adversarial attack, by marshalling and containing the attack, whilst maintaining the integrity of the overall system.

Thus, we feel that a hybrid configuration of models would offer considerable potential, balancing performance requirements without jeopardizing security efficacy. All application security controls can be embedded (Model B approach), whereas the network, transport, and session security can be incorporated in the DMZ (Model A approach). This segregation of security roles will offer the potential of delivering the users' requirements.
\section{Conclusions and Future Directions}
The performance of Model B is superior to that of Model A. This is because there is an additional overhead of a DMZ cloud in Model A with rule-based traffic forwarded through multiple hops of servers. Every hop added in between causes a delay that is reflected in the simulation results.

In contrast, there is only one hop between users and cloud servers in Model B, and there are no forwarding rules (because every server acts as a security node, as well).

For Model B, the performance is within the expected limits and there are no TCP delays, TCP segment delays and TCP re-transmissions. It is observed that the performance of database query response, HTML page and object responses is improved over Model A, and is satisfactory from the user’s perspective. The response times returned in this model are as expected from a cloud hosting of BI. Elimination of the UTM cloud layer has ensured that all of the session bottlenecks in the network are eliminated.

Hence, a combined model with application security embedded on the BI cloud, with the rest deployed in the UTM cloud, may be an optimum solution, both in terms of performance and effectiveness of security controls. In such a mixed model, the security controls will be technically sound with appropriate positioning of security layers, better security processes, clearly defined roles and accountabilities (application security and network security), and better effectiveness of controls.

Additional research is required to assess how such a hybrid model of BI security on the cloud can be implemented, and is the mandate for this research going forward. In particular, we are investigating how traditional security components (like those shown in the Model A DMZ) can be implemented in the form of cloud arrays in distributed architectures.

\end{document}